\documentclass[journal]{IEEEtran}
\usepackage{amsmath}
\usepackage{graphicx}
\usepackage{epstopdf} 
\usepackage{subfig}
\usepackage{url}
\usepackage{textcomp}
\graphicspath{ {figures/} }
\usepackage{textcomp}
\usepackage[utf8]{inputenc}
\usepackage{bbold}
\usepackage{amsmath}
\usepackage{multirow}
\usepackage{breqn}

\newcommand{\ket}[1]{| #1 \rangle}
\newcommand{\bra}[1]{\langle #1 |}

\title{A High Speed Integrated Quantum Random Number Generator with on-Chip Real-Time Randomness Extraction}

\author{\fontsize{6pt}{12pt} Francesco Regazzoni$^{1,2}$$^*$$^+$ \thanks{$^*$Corresponding Author: f.regazzoni@uva.nl, regazzoni@alari.ch}, Emna Amri$^{3,4}$$^+$\thanks{$^+$These two authors contributed equally}, Samuel Burri$^5$, Davide Rusca$^3$, Hugo Zbinden$^3$ and Edoardo Charbon$^5$\\
$^{1}$University of Amsterdam, Amsterdam (The Netherlands); $^{2}$ALaRI, Università della Svizzera italiana, Lugano (Switzerland); $^{3}$Group of Applied Physics (GAP), University of Geneva, Geneva (Switzerland); $^{4}$Id Quantique SA (IDQ), Geneva (Switzerland); $^{5}$École Polytechnique Fédérale de Lausanne (EPFL), Neuch\^{a}tel (Switzerland)
}

\begin{document}

\maketitle

\begin{abstract} 
The security of electronic devices has become a key requisite for the rapidly-expanding pervasive and hyper-connected world. Robust security protocols ensuring secure communication, device's resilience to attacks, authentication control and users privacy need to be implemented. Random Number Generators (RNGs) are the fundamental primitive in most secure protocols but, often, also the weakest one. Establishing security in billions of devices requires high quality random data generated at a sufficiently high throughput. On the other hand, the RNG should exhibit a high integration level with on-chip extraction to remove, in real time, potential imperfections. We present the first integrated Quantum RNG (QRNG) in a standard CMOS technology node. The QRNG is based on a parallel array of independent Single-Photon Avalanche Diodes (SPADs), homogeneously illuminated by a DC-biased LED, and co-integrated logic circuits for postprocessing. We describe the randomness generation process  and we prove the quantum origin of entropy. We show that  co-integration of combinational logic, even of high complexity, does not affect the quality of randomness. Our CMOS QRNG can reach up to 400 Mbit/s throughput with low power consumption.
Thanks to the use of standard CMOS technology and a modular architecture, our QRNG is suitable for a highly scalable solution.

\end{abstract}

\footnotetext{This work has been submitted to the IEEE for possible publication. Copyright may be transferred without notice, after which this version may no longer be accessible.}

With the ubiquity and density of mobile devices, security has become a very serious concern and, at the same time, a key enabler for next generation Internet-of-Things (IoT) and cyber-physical systems. True random number generators (TRNGs) are the core building block in almost all security schemes today \cite{sunar2009true,verbauwhede201524}. TRNGs exploit complex physical phenomena in order to generate randomness. However, their generation process is taken to be random without a complete physical model that could describe their behavior. This means that a possibly perfect entropy source could become completely deterministic with a higher understanding of the physical process. 
\begin{figure}
        \centering
        \includegraphics[trim={0cm 3cm 0cm 3cm},scale=0.35]{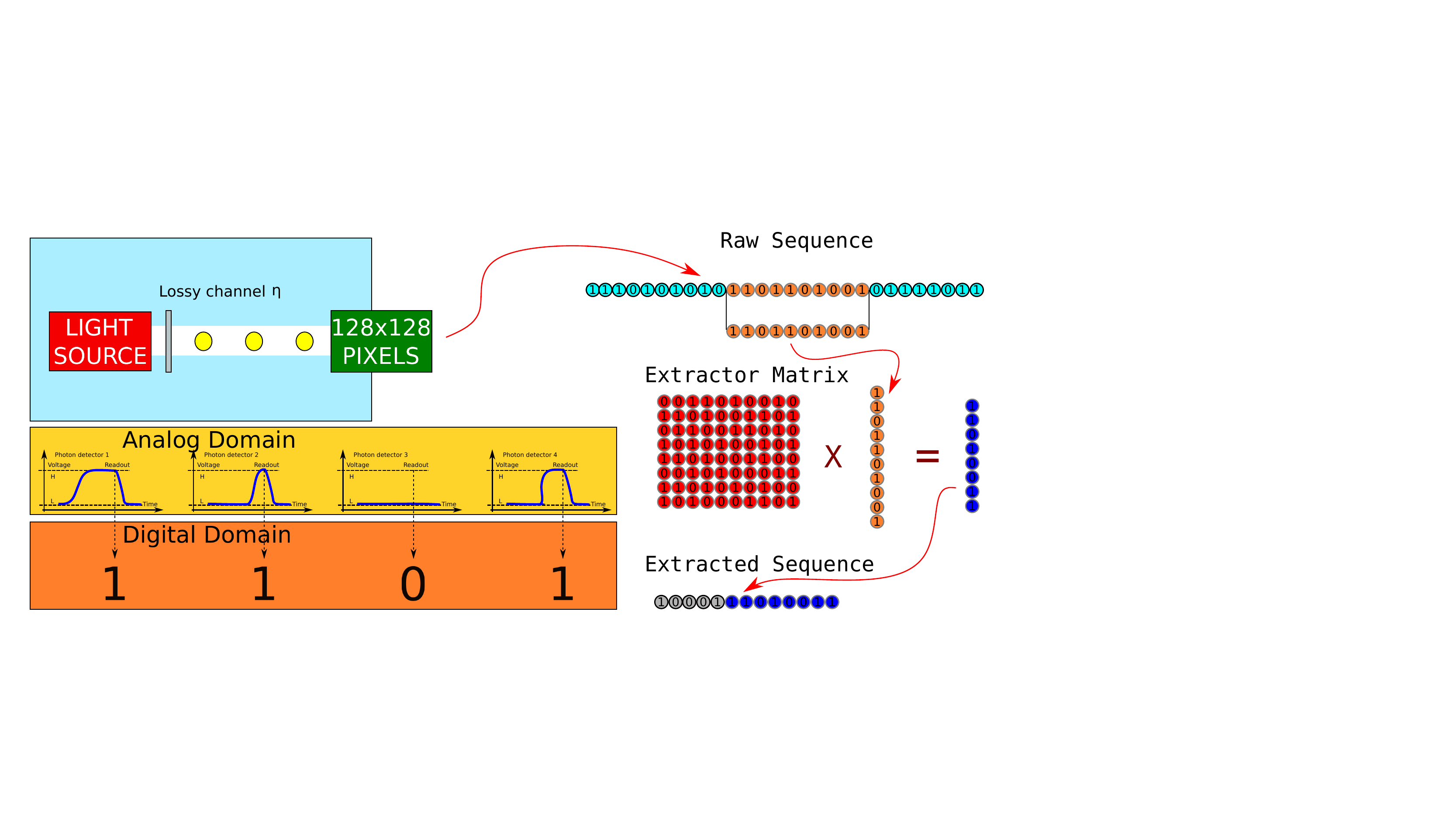}
        \caption{High level schematic illustration of the proposed QRNG and its operational principle. A light source illuminates a photon detector through a lossy channel with a transmission probability $\eta$ including all the optical coupling losses and the photon detection efficiency of the detector. The photon detector plays the role of the transducer and outputs digital bits correlated to the amount of detected photons. Raw data, that are directly accessible for testing purpose, are immediately forwarded into an entropy extractor, also implemented on chip, that produces the sequence of true random numbers.}
	    \label{fig:qrngoverview}    
\end{figure}
QRNGs solve this problem by exploiting the intrinsic probabilistic nature of quantum mechanics. In fact, the measurement result of quantum processes can not be predicted, not even by an all powerful malicious third party. Several implementations have been proposed, while the specific requirements are depending on target applications. Numerous works focus on extremely high-speed architectures~\cite{wei2011high}, some on extracting entropy from existing devices such as FPGAs~\cite{rovzic2016iterating} or general-purpose processors~\cite{mechalas2014intel} and others on using optical entropy sources~\cite{massari201616}. Our QRNG lies within the last category and can be modelled at the first level as a light source emitting photons with a certain statistical time distribution and illuminating a photon detector through a lossy channel with a transmission probability $\eta$ including all the optical coupling losses and the photon detection efficiency of the detector. The photon detector plays the role of the transducer and outputs digital bits correlated to the amount of detected photons. Due to hardware imperfections and external classical noise, the quantum randomness is inevitably mixed with classical randomness. To distill the randomness of quantum origin, one needs to quantify it using an appropriate physical modelling and apply a specialized mathematical post-processing called randomness extractor. Figure ~\ref{fig:qrngoverview} shows a demonstration of this model.

In this paper, we present an integrated QRNG with on-chip real-time randomness extraction. The QRNG is based on a parallel array of independent CMOS Single-Photon Avalanche Diodes (SPADs), homogeneously illuminated by a DC-biased LED.  Digital post-processing is implemented on chip and provides a true random bit-stream through two parallel digital interfaces. Our device is suitable for applications requiring an integrated architecture, capable of guaranteeing the required throughput in real time~\cite{verbauwhede201524}. 
We describe the Integrated Circuit (IC) architecture and we discuss the randomness generation process and the quantum entropy extraction. We propose a model to ensure that the produced randomness is indeed originated from a quantum phenomenon and we evaluate the influence that the co-integrated digital post-processing may have on the raw randomness quality, before performing statistical analysis on data after randomness extraction. Our CMOS QRNG can output up to 400 Mbit/s of data with significant area reduction and low energy per generated bit.

\section{QRNG with Integrated Entropy Extraction Architecture}

\begin{figure}
    \centering\includegraphics[scale=0.3]{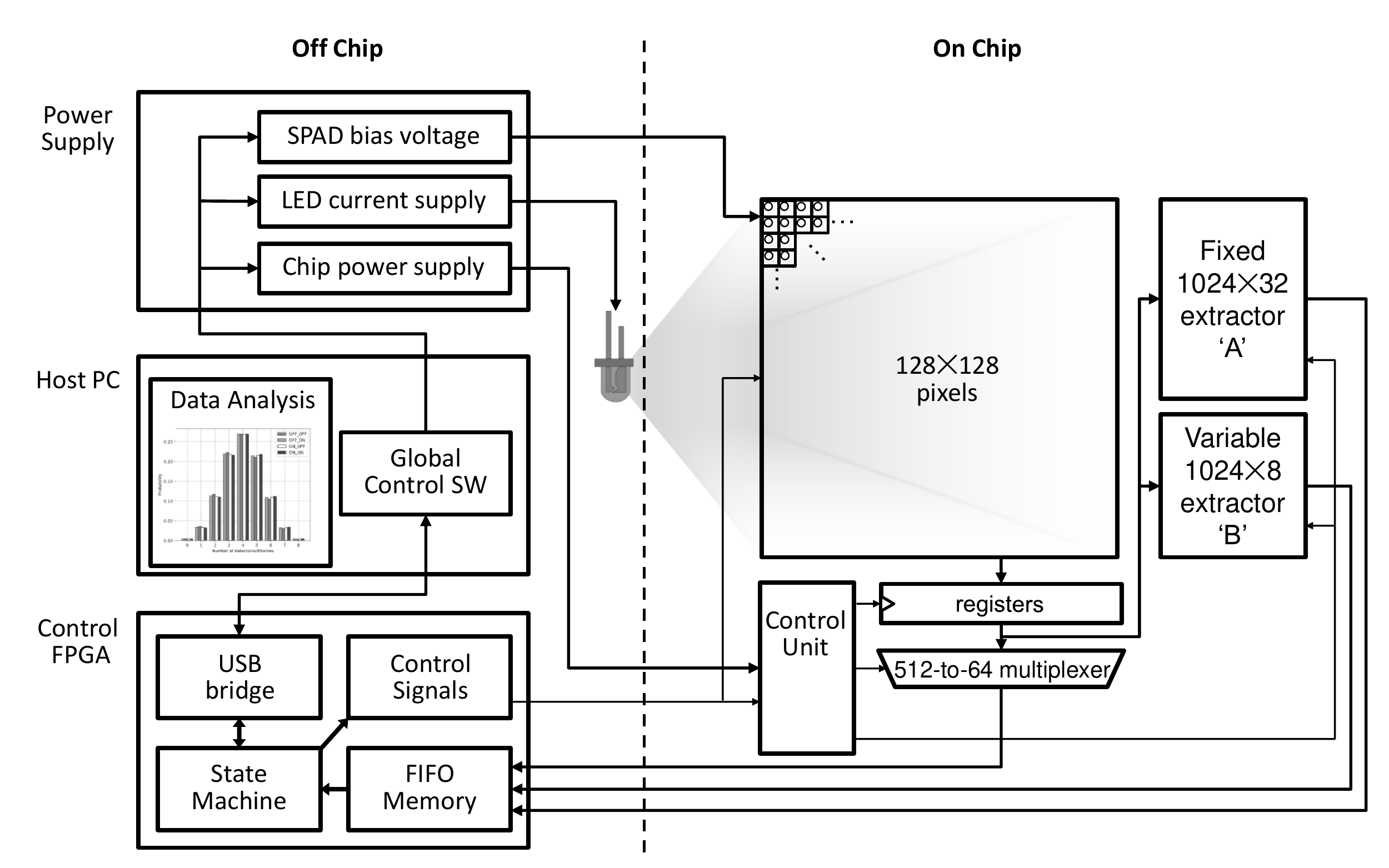}
    \caption{Overview of the experimental setup architecture: It comprises a LED source, the QRNG integrated chip, and a custom made FPGA used to control the QRNG chip and to format and transmit the bit stream to the host PC where the sequence of data are analyzed.}
    \label{fig:system-overview}
\end{figure}

Our system is depicted in Figure~\ref{fig:system-overview}; it features a QRNG IC implementing the SPADs matrix, two extractors, and a high-speed digital interface. The entropy source is obtained from a LED illuminating an array of 16,384 single-photon avalanche diodes (SPADs), which convert photons  Poissionian distributed in time, onto a 2D constellation of binary digits '1' if at least a photon is detected and '0' otherwise. The chip features two extractors based on vector-matrix multiplication to generate high entropy bit sequences starting from the original constellation. The two extractors target different application requirements. The first is a fixed extractor, denominated ‘A’; it is realized from standard logic gates using a hard-coded n x k matrix pre-generated from an independent QRNG; this extractor is intended for applications requiring higher throughput and minimal area. The second is a variable extractor, denominated ‘B’; it is a matrix of memory elements, realized by means of scan chain registers to minimize the amount of pads needed for programming it and to allow the use of design tools; this extractor is intended for applications requiring one to extract an ad hoc matrix in the field for a higher security level. We implemented both extractors on this proof-of-principle chip in order to demonstrate the feasibility of the co-integration with both extraction methods.\\
The QRNG IC was fabricated using a technology where the SPADs exhibit low noise, negligible afterpulsing, and low crosstalk. These characteristics are essential to enable the maximum speed with the desired randomness quality.

\begin{figure}[h!]
    \centering\includegraphics[scale=0.35]{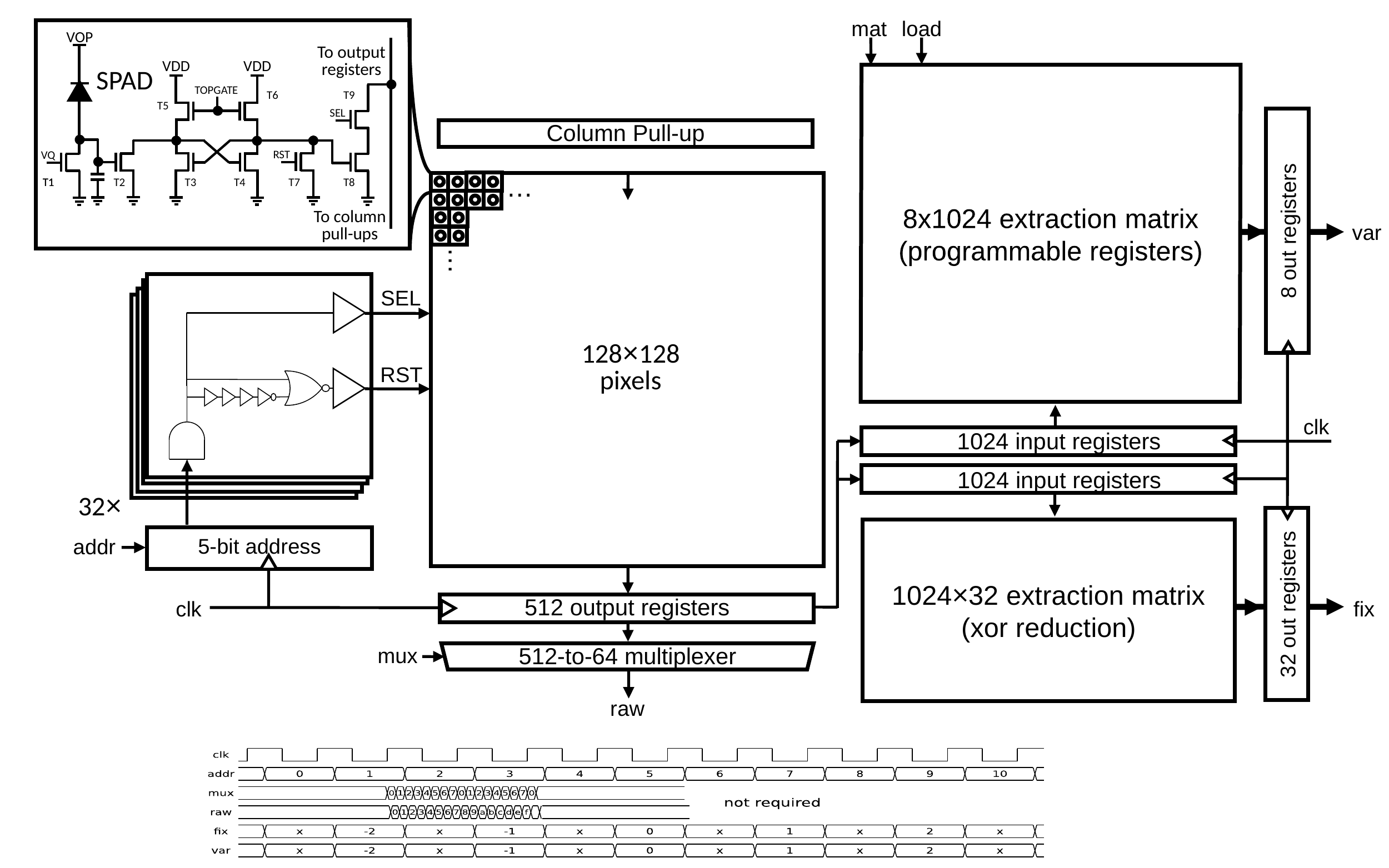}
    \caption{Block and timing diagram of the proposed QRNG chip. The chip comprises an array of 16,384 independent SPADs illuminated through a diffuser made of a standard 4$\mu$m polyimide layer. The SPADs act as entropy sources that generate a fast bit stream, which is then transformed onto a fully randomized 400Mbit/s binary stream via an extractor
integrated on chip. The timing diagram of the readout process is shown in the inset.}
    \label{fig:block-diagram}
\end{figure}

\begin{figure}[h!]
    {\includegraphics[trim={5cm 5cm 0 6cm},scale=0.4]{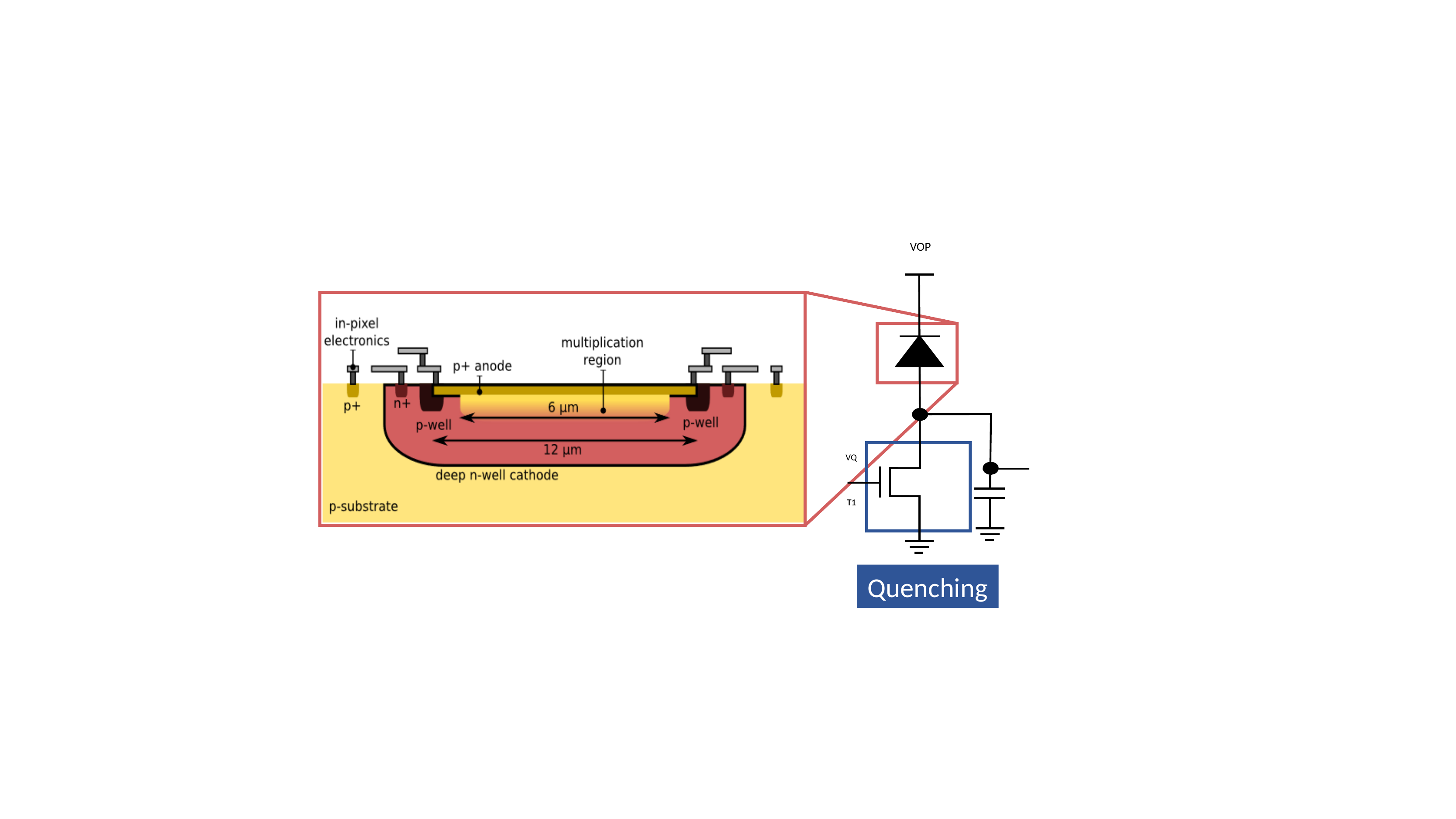} }%
    \caption{One pixel's architecture. The quenching circuit on the right. A scheme of the SPAD cross-section on the left.}%
    \label{spad-details}%
\end{figure}

\begin{figure}[h!]
    {\includegraphics[scale=0.4]{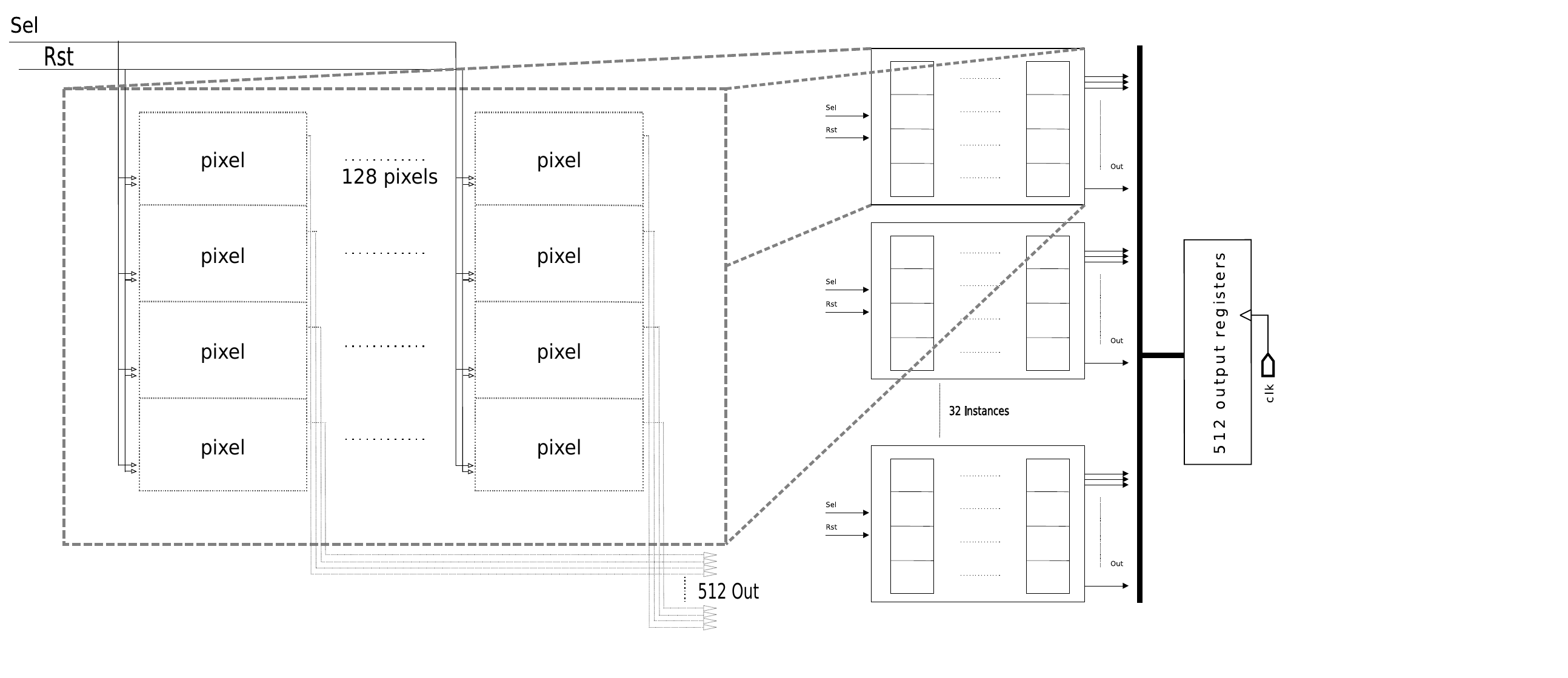} }%
    \caption{Pixels connections and Readout. The 512 pixels belonging to the same group of 4 rows are connected to the same \emph{Sel} and \emph{Rst} but are connected to 512 independent outputs. Each of the 32 groups 4 lines has independent \emph{Sel} and \emph{Rst} signal, but they are sharing, using a multiplexer, the same 512 output register.}%
    \label{spad-details-schem}%
\end{figure}

The block diagram of the QRNG is shown in Figure~\ref{fig:block-diagram}. The chip comprises an array of SPADs organized in a matrix of 128x128 pixels, whose schematic is shown in the inset of the figure. The architecture of the pixel is depicted in Figure~\ref{spad-details}. In each pixel, a SPAD implemented as a P+/N-well junction with lightly doped N wells as guard rings, is passively quenched and recharged by means of the transistor T1. Transistor T2 is used to trigger the static embedded all-NMOS memory (T3, T4, T5, T6), while T7 is used as reset, and T8 and T9 are used to read out the content of the memory in a random access fashion. The matrix is read out four-rows-in-parallel every 40~ns and the raw data is stored in a 512-bit register whose content is diverted to the two extractors. All the pixels belonging to the same group of four rows are connected to the same reset (\emph{rst}) and the same selection signal (\emph{sel}), as depicted in Figure~\ref{spad-details-schem}. The entire matrix is read out in 1.3$~\mu$s.

\begin{figure}[h!]
    {\includegraphics[trim={8cm 0cm 0 0cm},scale=0.13]{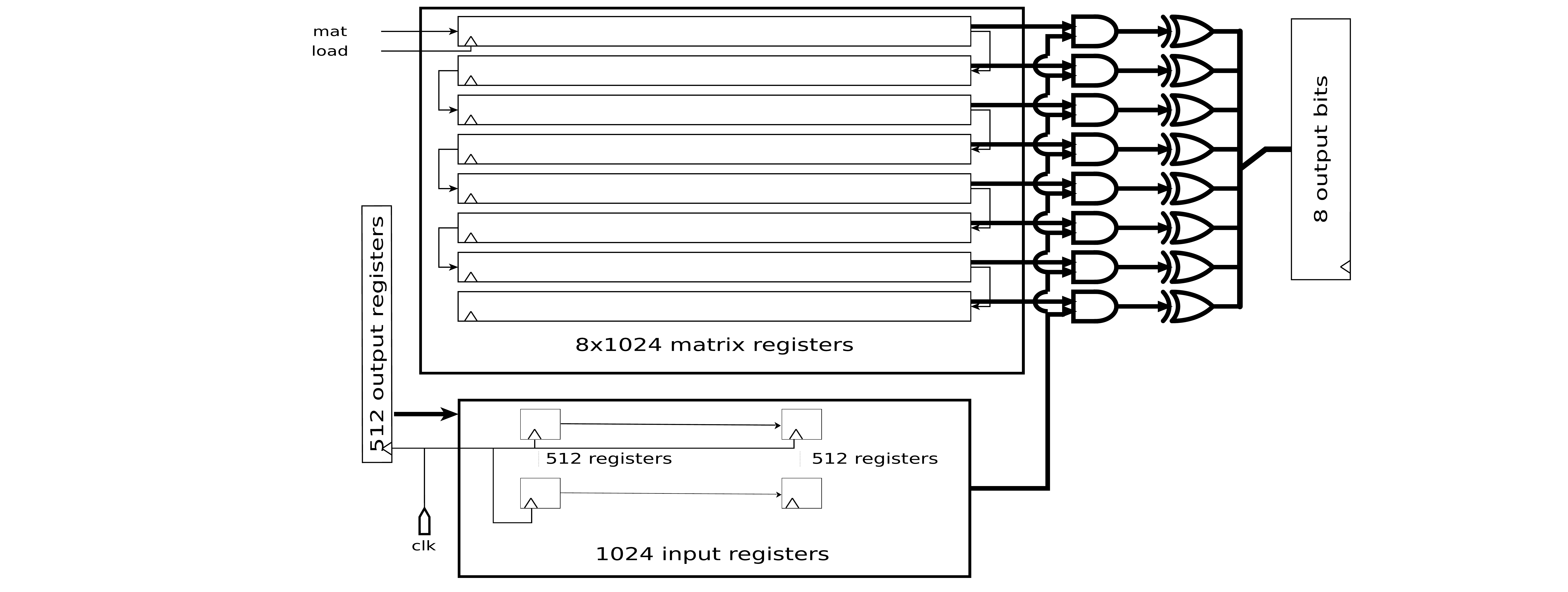} }%
    \caption{Extraction using Extractor 'B'. The 512 entropy bits at the input are feed the first group of 512 registers. When a input is ready, the data in the first group of registers are shifted into the second group. Once all the 1024 bit are filled with new data, they are ANDed with each line of the matrix and XORed to produce the output.}%
    \label{extractorb}%
\end{figure}

Extractor ‘A’ comprises 1024x32 XOR reduction cells fed through an intermediate 1024-bit register. Extractor ‘B’, depicted in Figure~\ref{extractorb} is an array of 8x1024 SRAM cells whose content is provided externally. Both extractors are operating within the chip in real time: extractor ‘A’ can reach a throughput of 400~Mbit/s, extractor ‘B’ of 100~Mbit/s, both producing random words at 12.5MHz. The resulting bit streams are read out from the chip in 8-bit and 32-bit packets, using two parallel interfaces. In test mode, we can independently access the raw entropy source output at 1.6~Gbit/s to verify the quality of raw data at the source in accordance with high security standards~\cite{niststandards}. 

\begin{figure*}%
    \centering
    \subfloat[~]{{\includegraphics[scale=0.35]{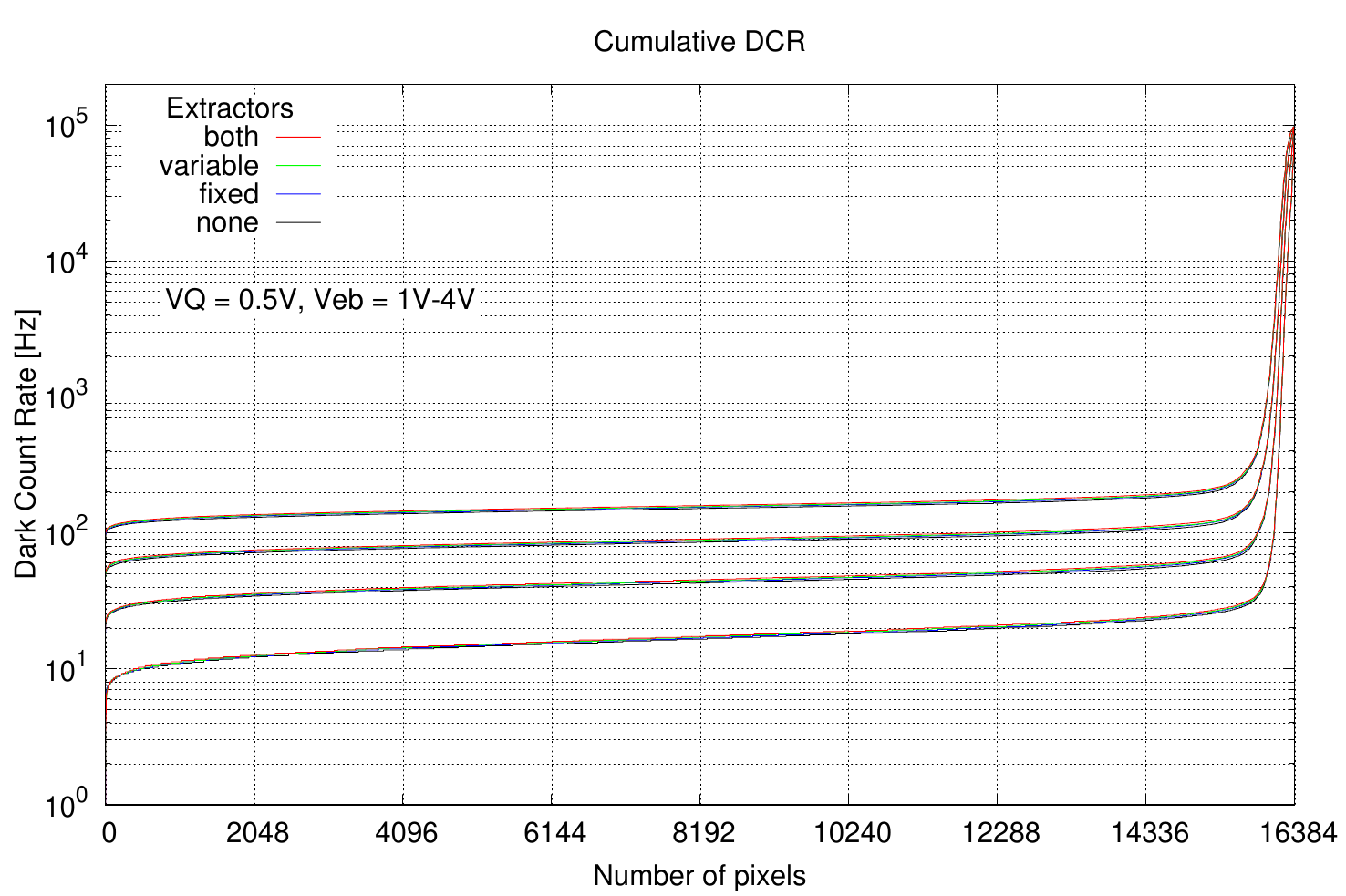}}}%
    \qquad
    \subfloat[~]{{\includegraphics[scale=0.35]{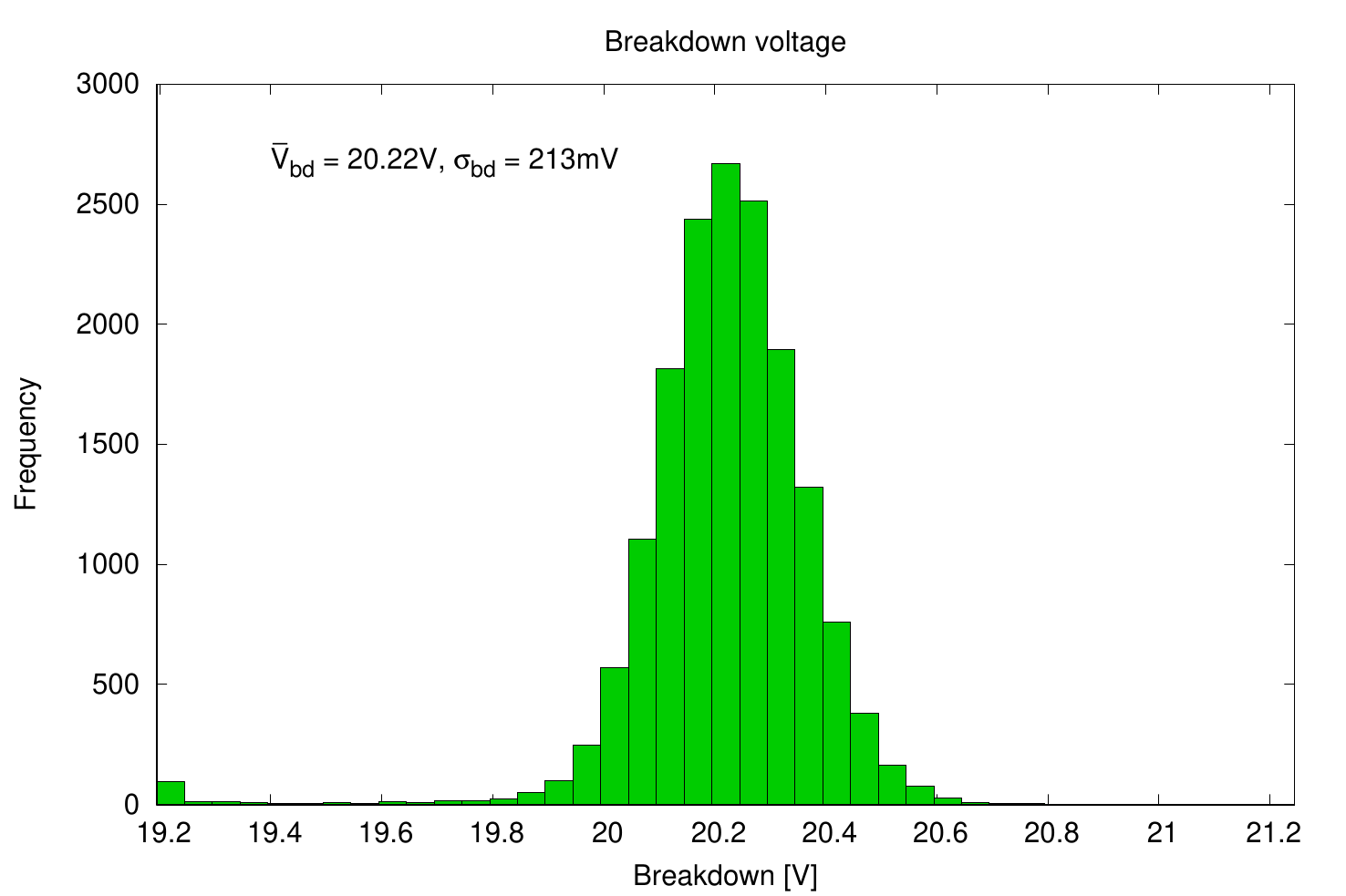} }}\\%
    \qquad
    \subfloat[~]{{\includegraphics[scale=0.35]{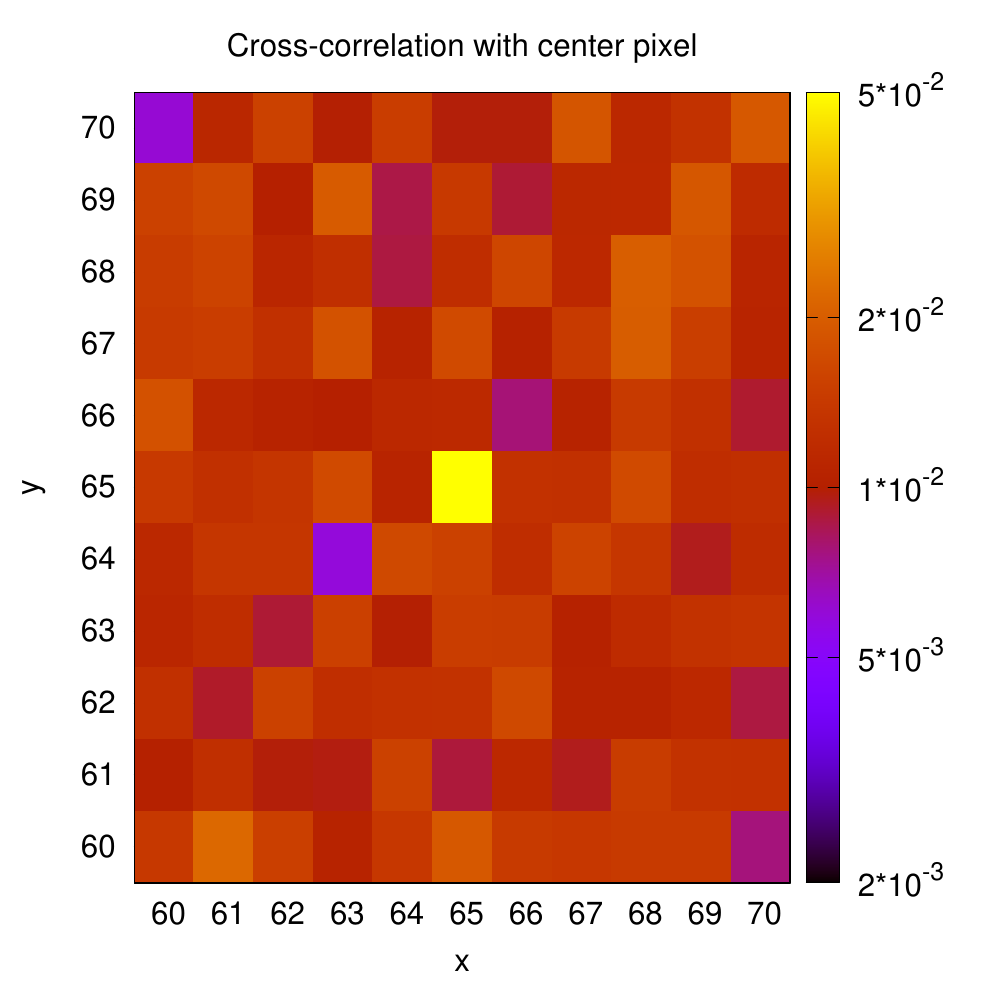} }}%
    \qquad
    \subfloat[~]{{\includegraphics[scale=0.35]{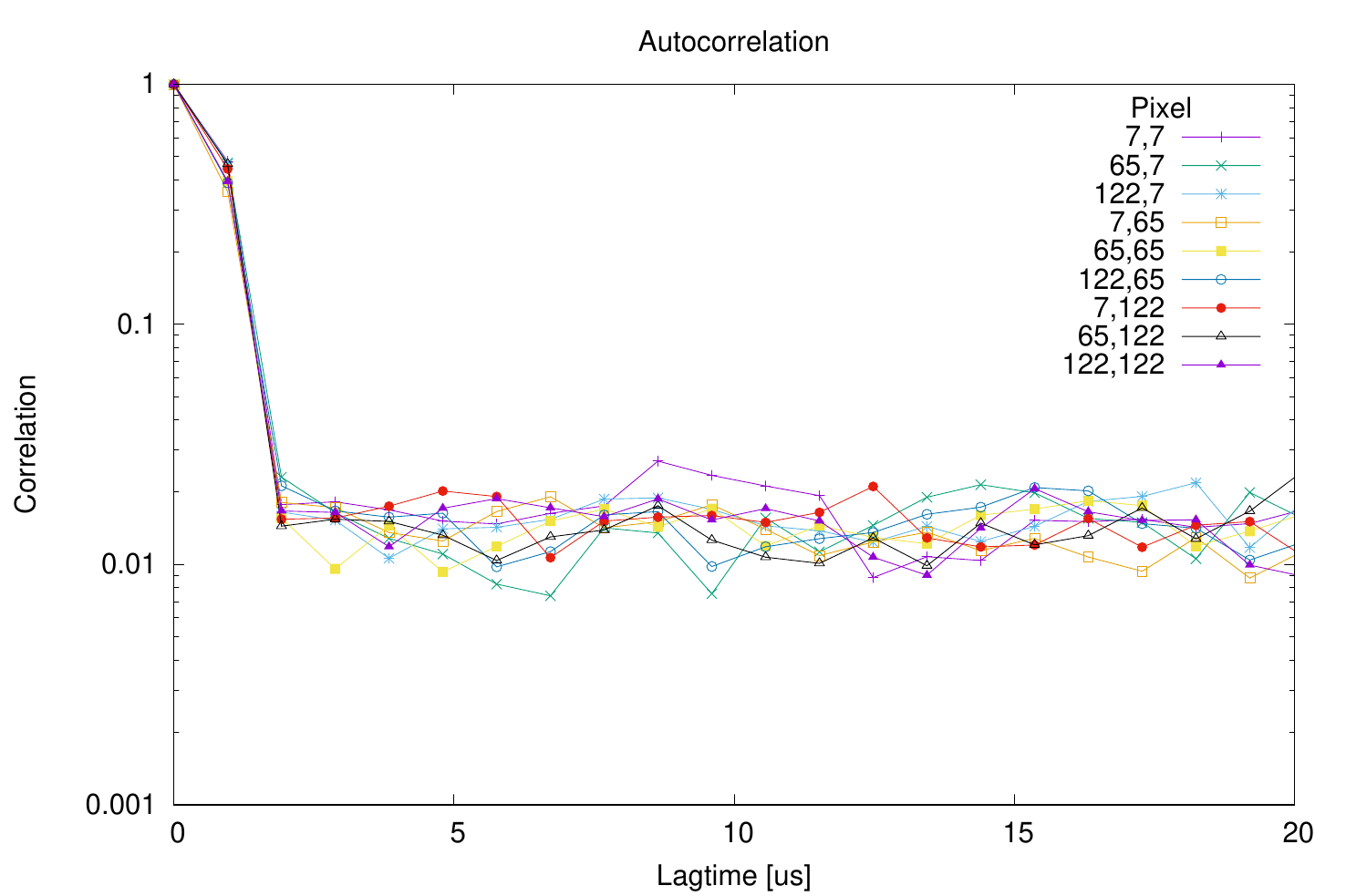} }}%
    
    \caption{Performance of the SPAD used in this work. Clockwise from top-left: (a) Cumulative dark count rate for different excess bias and under different operation conditions (Extractors ON and/or OFF). Note that the 512 noisy pixels called "hot pixels" are taken into account in the entropy model; (b) Breakdown voltage distribution; (d) Autocorrelation function for several pixels vs. lag time; (c) Cross-correlation function with respect to surrounding 11 x 11 pixels. All measurements at room temperature.
}%
\label{spad-performance}%
\end{figure*}

The SPADs characterization results are gathered in Figure~\ref{spad-performance}. The figure shows a plot of the breakdown voltage distribution and of the dark count rate (DCR) across the entire population of pixels. The afterpulsing probability (APP) was characterized by means of the autocorrelation of several pixels in time, based on their digital output. Crosstalk was measured as the cross-correlation of a central pixel with all the surrounding pixels. The plot in the figure shows central pixel (7,7) within a surrounding 11x11 pixel matrix. The excess bias voltage and dead time used in this chip were 2.1~V and 1.3$~\mu$s, respectively, so as to keep APP and crosstalk below 0.1\%.

\section{Effects of Integration on Entropy} 

\begin{figure}%
    \centering
    \subfloat[~]{{\includegraphics[scale=0.25]{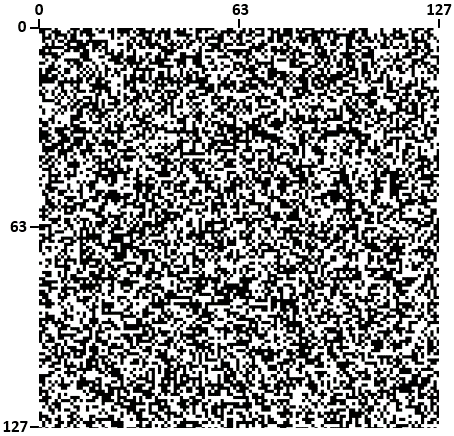} }}%
    \qquad
    \subfloat[~]{{\includegraphics[scale=0.25]{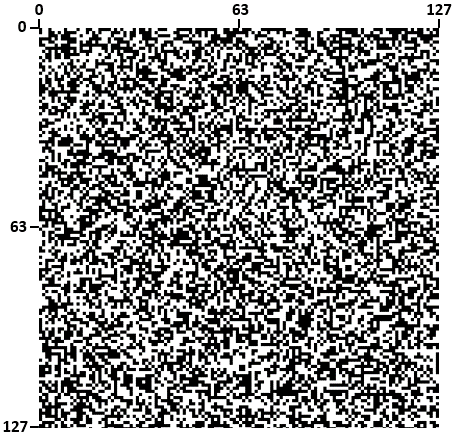} }}%
    \\
    \subfloat[~]{{\includegraphics[scale=0.25]{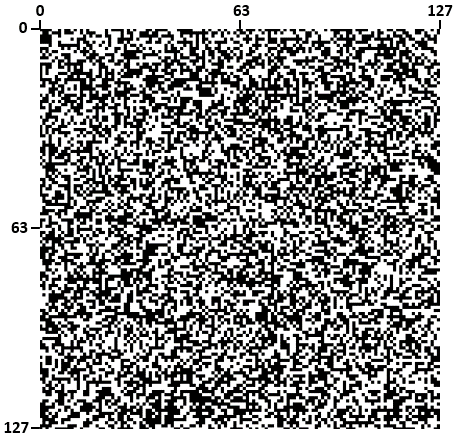} }}%
    \qquad
    \subfloat[~]{{\includegraphics[scale=0.25]{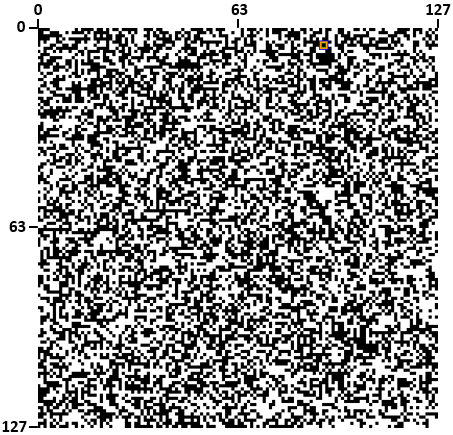} }}%

    \caption{Performance of the QRNG when different extractors are active. Raw binary constellations. (a) Both extractors are OFF; (b) the fixed extractor is ON and the variable extractor is OFF; (c) the fixed extractor is OFF and the variable extractor is ON; (d) both extractors are ON. }%
    \label{randomness-extractor}%
\end{figure}

\begin{figure}
    \centering
    {\includegraphics[trim={0cm 0cm 0cm 1cm},scale=0.5]{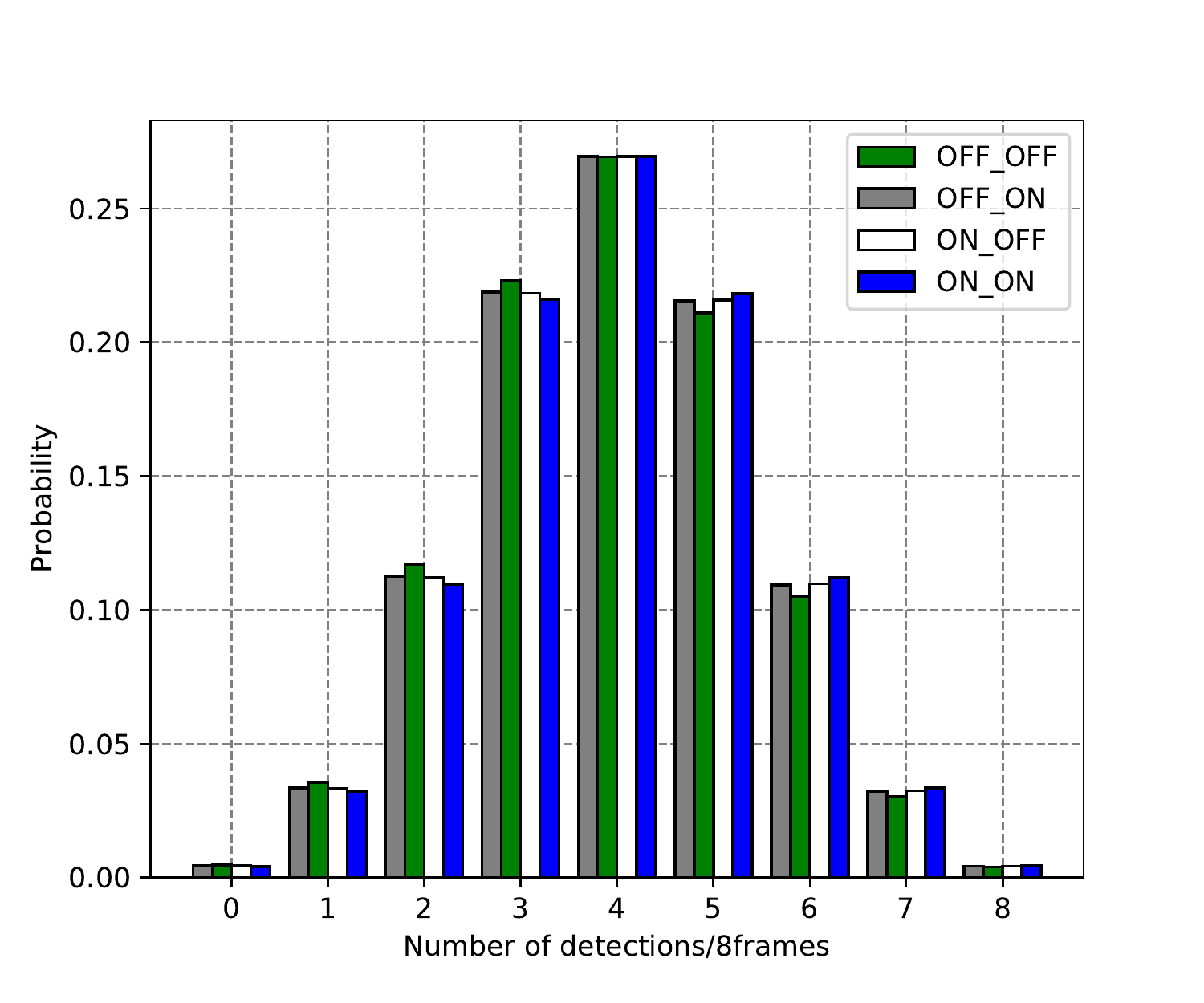} }
    \caption{Detection histogram. Detection histogram (8-bits Hamming Weight distributions) of raw data for different operating conditions: (green) Both extractors are OFF; (grey) the fixed extractor is OFF and the variable extractor is ON; (white) the fixed extractor is ON and the variable extractor is OFF; (blue) both extractors are ON.}%
    \label{fig:histogram}%
\end{figure}

Our first concern was the effect of electromagnetic interference introduced on the entropy source by the neighboring circuitry used for on-chip digital post-processing. 
That is why, we started by evaluating the influence of on-chip extractors on the quality of the generated randomness. This information is crucial to demonstrate the feasibility of a fully integrated QRNG in a commercially relevant CMOS technology. We compared the statistical distribution of raw data (before post-processing) when the two extractors are not powered on, when only one of them is working and when both are performing on-the-fly extraction for four QRNG chips. Figure~\ref{randomness-extractor} shows the constellations generated by the SPAD array for these same conditions. From the figure the equi-probability of ‘1’s and ‘0’s is apparent (and actually we computed a probability of 50.36\% of '1' vs 49.64\% of '0') with a homogeneous distribution among the pixels, irrespective of the state of the extractors, at better than 2$\sigma$ deviation from the 50\% distribution mark. This observation was confirmed when we compared the detection histograms (8-bits hamming weight distributions) in all four cases (shown in Figure~\ref{fig:histogram}). Therefore, we can conclusively confirm that the analog-digital co-integration doesn't have any quality-deterioration effect on the quantum process generation and measurement.

\begin{figure}
    \centering\includegraphics[scale=0.4]{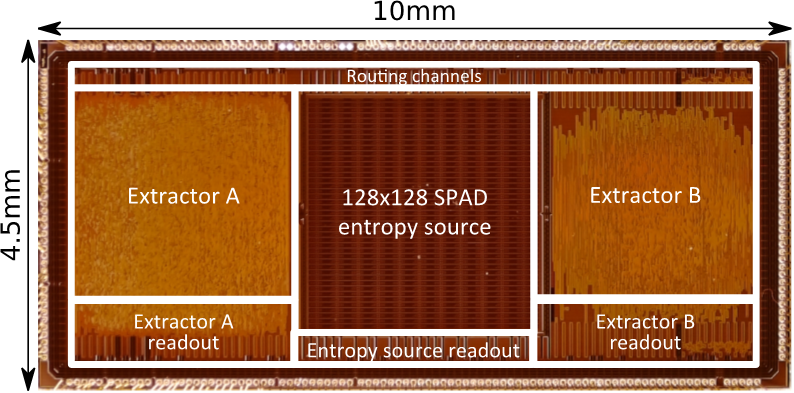}
    \caption{Chip micrograph. Complete chip micrograph of the proposed QRNG chip implemented in standard 0.35$\mu$m CMOS technology. From left, it is possible to see the Extractor A and its read out at the bottom, the entropy source composed by a matrix of 128$\times$128 SPADs and its readout at the bottom, the Extractor B and its readout at the bottom.}
    \label{fig:chip-micrograph}
\end{figure}

\section{Modeling the Quantum Randomness}

Before evaluating the performance of our device as a quantum random number generator, we  need to ensure that the produced randomness is due to a quantum process (and not coming , for instance, from classical parasitic noise). We propose a model to demonstrate that, under given assumptions of the possible knowledge of an adversary, the randomness produced by our QRNG is indeed quantum.

In the QRNG that we propose here, the entropy comes from an LED that illuminates a matrix of SPADs. The quantum process that is exploited is the distribution of photons over the surface of the detectors. The position of each photon is not deterministically known before its detection. The probability distribution of its position is given by the intensity profile of the electromagnetic field impinging on the detectors. The distance between the detectors and the source is chosen such that the light intensity profile on the detectors is uniform. Considering the time interval defined by the acquisition time and the space mode spanned by the area of the detectors matrix (spatially we consider the state in a single mode), the quantum state emitted by the LED is a mixed state in the Fock space, where the probability of having $n$ photons is given by the Poisson distribution $P_N$:

\begin{equation}\label{Poisson}
 P_N(n)=e^{-\lambda}\frac{\lambda^{n}}{n!}.
 \end{equation}

We model the matrix of SPADs as a collection of independent detectors. Their efficiency $\eta$ is considered in our model, as the probability that one photon is detected. If $n$ photons arrive on one detector, the probability that at least one of them is detected (probability that the detector clicks) is given by $P_{det}^n = 1-(1-\eta)^n$. Similarly, we can model the dark counts probability with a random variable $S_i = 0,1$ for each detector. In the case of $S_i = 1$ the $i^{th}$ detector will click independently of the light coming into it. This event has probability $p_{dark}$ equal to the probability of having a dark count on one pixel.

Using the methods proposed by Frauchiger \textit{et al}.~\cite{Daniela} to evaluate the amount of quantum randomness of our device we evaluate the min-Entropy of the output distribution conditioned on the distribution of all the predictable side information.
In our case, the side information is modeled as classical and determined by the random variable $E$. The conditional min-Entropy that we need to compute is thus the following:

\begin{equation}
    H_{min}(X|E) = -\sum_{e} P_E(e) \log_2 [\max_x P_{X|E=e}(x|e)],
\end{equation}
where $X$ represents the random variable of the output sequence, $P_{X|E}$ is the conditional probability distribution of $X$ knowing the variable $E$, and the quantity $2^{-H_{min}(X|E)}$ represents the maximum guessing probability of $X$ given $E$. In order to find the probability distribution of the output sequence we have to characterize the quantum state of our device $\rho$, and the measurement represented by the operators $\Pi^{\overline{x}}$, where $\overline{x}$ is a possible output bit string. Moreover, all side information is represented by the value $e$ which is the result of the measurement $E^{e}$ that represent the part of the process considered to be deterministic. Provided the physical characterization of the device, the probability distribution is given by the Born rule:

\begin{equation}
    P_{\overline{X},E}(\overline{x},e) = Tr[\Pi^{\overline{x}}E^e\rho(E^e)^\dag]
\end{equation}
without any need for a stochastic model (see Appendix~\ref{app:model} for more information). The most critical part of a QRNG is however defining all the possible side information. In our case, the photon distribution of the source and the random variables corresponding to the dark counts are the principal sources of classical side information and are considered as completely foreseeable. The event of a photon to be detected (described by the efficiency $\eta$) is considered to be independent of any possible observer. However, in this conservative model, we consider a possible shot to shot uncertainty with respect to the average observed value of the efficiency.

All the possible sources of side information have been characterized and the following parameters have been measured accurately in order to certify that the extracted randomness is of quantum nature:  $\eta = 0.12 \pm 0.03$ (considering a possible deviation of 25\% from the average efficiency), $p_{dark} = 8.45\cdot10^{-5}$ per detection window, to this probability we added also the probability of cross-talk of around $P_{cross} = 0.001$ (this corresponds to the worst case scenario where each click provoked by the click of another detector is known by an adversary) and the mean photon number $\lambda$ of photons arriving on the detectors chosen in such a way that the probability of each of them to click is equal to $0.5$ (as set experimentally). Moreover in the post-processing we took into account the 512 hot pixels which are always on, this accounts for an additional lowering of the entropy. With this characterization of the device, the min-Entropy per bit has been estimated to be $H_{min}(X|E)/m \approx 73\% $, which fixes a limit on the possible extractable randomness from the raw generated sequence of bits.

In this work an extractor based on vector-matrix multiplication~\cite{IDQ} was used. This extractor is applied to vectors of $n$ raw bits and output shorter vectors of $k$ random bits. Once the length of the raw string $n$ is chosen ( in our case $n$ is equal to 1024), the parameter $k$ will be bound by the probability that the output bit string deviates from perfectly random output bits $\epsilon$. The extraction function used in our QRNG is taken from a two-universal family of hash functions, so it is possible to quantify this failure probability by the Leftover Hash Lemma with side information:

\begin{equation}\label{epsilon}
 \epsilon=2^{-(H_{min}(X|E)\cdot n-k)/2}.
\end{equation}

For instance with the obtained value of min-Entropy and for a vector of 1024 raw bits, the parameter k should be lower than 548 for efficient entropy extraction ($\epsilon=2^{-100}$). In our case the parameter k is equal to 32 for the fixed extractor and equal to 8 for the variable extractor which confirms the high efficiency of our on-chip extraction. Note that these values of $k$ were chosen according to a trade-off between a sufficiently high throughput and a minimum area. 

\section{Experimental methods}

\subsection{Fabrication of the Integrated QRNG}
The QRNG integrated circuit was fabricated in a standard 0.35$\mu$m 1P4M high-voltage CMOS technology through Europractice multi-project wafer foundry service. The SPAD device junction is formed between a p+ anode and deep n-well cathode embedded in the standard p-type substrate. A p-well guard ring is used to realize a uniform electric field and prevent premature edge-breakdown of the device. The SPAD is of circular shape with an active diameter of 6$\mu$m.

\subsection{Measurements}
For measurements the IC was bonded in a ceramic PGA-256 package and inserted in a socket on a adapter PCB for connection with the FPGA main board. The FPGA main board contains a Xilinx Spartan~6 FPGA connecting directly to the QRNG I/O and a Cypress FX3 USB transceiver for communication with a control computer. Voltage supplies for the chip are provided through the main board with separate jumpers for the extractor supply.

A Rohde\&Schwarz HMP2030 dc power supply was used to provide controllable operating bias and quenching voltage for the SPAD and pixel circuit. A custom-built matrix of relay switches (Kemet EC2-12TNU) and a USB-SMU (Agilent/Keysight U2723A) were used to respectively switch power of the extractor matrices and control the LED illumination current. For measurements in a temperature-controlled environment the QRNG together with the FPGA board was put in an ESPEC SU-262 temperature control chamber. The temperature chamber is outfitted with a nitrogen purge to avoid condensation.

To evaluate the breakdown voltage of the individual detectors the voltage was swept in steps of 50~mV from a voltage where no events are detected to where all detectors show activity. Through readout of individual detectors, the number of events during one second at each voltage step has been recorded and fitted with a two-piecewise linear function. The function evaluates to zero below the breakdown voltage and increases thereafter, thus providing the breakdown voltage for each detector.

Data from the extraction matrices is acquired through the control FPGA in the same way as the raw data needed to evaluate the breakdown voltage. The operating conditions, namely excess bias and quenching voltage and LED illumination current are set. Then synchronous control signals are applied to read out the data from the output register of the matrices. The control FPGA relays the data to a computer where they are stored on disk for analysis.

\subsection{Randomness Evaluation}
The random data sequences were stored off-chip and tested using the NIST Statistical Test Suite (SP 800-22) downloaded from the NIST website (\url{https://csrc.nist.gov/Projects/Random-Bit-Generation/Documentation-and-Software}). The parameters used to perform these tests are the followings: blockFrequencyBlockLength = 12800, nonOverlappingTemplateBlockLength = 10, overlappingTemplateBlockLength = 9, approximateEntropyBlockLength = 10, serialBlockLength = 16 and linearComplexitySequenceLength = 1000. The results are stored in an dedicated file, automatically generated when the tests are launched. 
The same data sequences were tested using the Diehard battery of tests available 
\url{https://webhome.phy.duke.edu/~rgb/General/dieharder.php}

\section{Results}

\begin{figure}[h!]
    {\includegraphics[trim={0cm 2cm 0cm 2.5cm},scale=0.35]{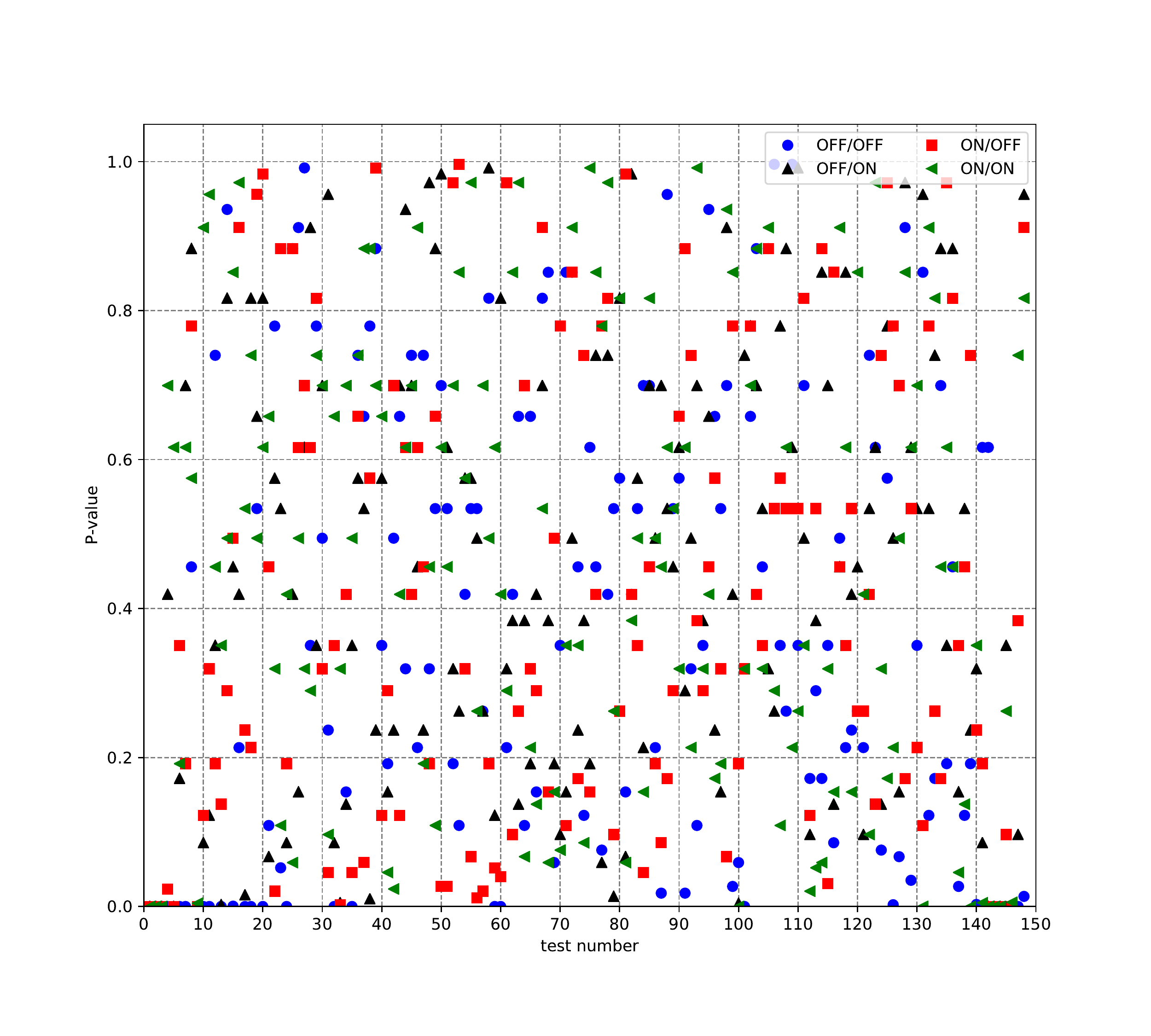} }%
    \caption{QRNG statistical characterization. The figure shows the P-value distribution for 145 Non-overlapping Template Matching (NIST test) done on raw data, provided by 4 chips. The purpose of the test is to detect the presence of a given (non-periodic) pattern in a 1-bit sliding window. We can see that p-values produced by this test are perfectly distributed between ‘0’ and ‘1’ with very few ‘0’ and 0.99 points.}%
    \label{qrng-performance}%
\end{figure}

First, the effect of temperature on the QRNG was evaluated and we saw a slight decrease of the quantum entropy per generated bit when increasing temperature. This decrease is due to a higher thermally induced noise that can lead to a bias toward ‘1’. Afterwards, about 1 Gbit of extracted data using both extractors, were tested using the NIST statistical test suite~\cite{bassham2010sp} and the DIEHARD test battery~\cite{marsaglia2008marsaglia} and it passed all of them. The NIST SP 800-22 tests were performed using 1000 sequences of 1 Mbit each. The minimum pass rate for each statistical test with the exception of the random excursion (variant) test is approximately = 980 for a sample size = 1000 binary sequences. The minimum pass rate for the random excursion (variant) test is approximately = 599 for a sample size = 613 binary sequences. The tests results for data generated by the fixed and the variable extractor are shown in Table~\ref{tab:nist_fix} and Table~\ref{tab:nist_var} respectively. The Diehard tests were performed on two data sets of 500 Mbits generated by the fixed extractor (P-Value 'A') and the variable extractor (P-value 'B'). The results are presented in Table~\ref{tab:diehard}. Finally,  Figure~\ref{qrng-performance} shows the P-value distribution for 145 Non-overlapping Template Matching (NIST test) done on raw data on 4 different chips.

\begin{table}
\centering
\small
\caption{NIST tests results: The uniformity of P-values and the proportion of passing sequences for 1000 samples of 1 Mbit each generated by the fixed extractor 'A'}

\begin{tabular}{lcc}
\hline
Statistical test & P-Value & Proportion \\
\hline
Frequency & 0.99577 & 988/1000 \\
Block-Frequency & 0.239266  &  990/1000\\
CumulativeSums1 & 0.941144  &  989/1000\\
CumulativeSums2 & 0.062427  &  986/1000 \\
Runs & 0.928857 &  989/1000\\
LongestRun & 0.473064  &  993/1000\\
Rank & 0.301194  &  990/1000\\
FFT & 0.029205  &  983/1000\\
NonOverlappingTemplate  & 0.662091 & 993/1000 \\
OverlappingTemplate & 0.735908  &  988/1000  \\
Universal & 0.829047 &   991/1000 \\
ApproximateEntropy & 0.767582  &  988/1000  \\
RandomExcursions & 0.349160 &   606/613 \\
RandomExcursionsVariant & 0.359855  &  607/613\\
Serial & 0.800005  &  994/1000 \\
LinearComplexity & 0.747898  &  994/1000  \\
\hline
\end{tabular}
\label{tab:nist_fix}
\end{table}

\begin{table}
\centering
\small
\caption{NIST tests results: The uniformity of P-values and the proportion of passing sequences for 1000 samples of 1 Mbit each generated by the variable extractor 'B'}
\begin{tabular}{lcc}
\hline
Statistical test   & P-Value & Proportion\\
\hline
Frequency & 0.614226   & 991/1000 \\
Block-Frequency & 0.616305  &  995/1000\\
CumulativeSums1 & 0.626709  &  990/1000 \\
CumulativeSums2 & 0.980341  &  994/1000 \\
Runs & 0.190654  &  989/1000 \\
LongestRun & 0.612147  &  986/1000\\
Rank & 0.452173  &  988/1000 \\
FFT & 0.870856  &  989/1000\\
NonOverlappingTemplate & 0.757790  &  993/1000 \\
OverlappingTemplate & 0.039073   & 988/1000  \\
Universal & 0.014550  &  988/1000 \\
ApproximateEntropy & 0.221317  &  987/1000 \\
RandomExcursions & 0.822122  &  607/613 \\
RandomExcursionsVariant & 0.367493 &   610/613 \\
Serial & 0.777265  &  994/1000 \\
LinearComplexity & 0.809249  &  992/1000 \\
\hline
\end{tabular}
\label{tab:nist_var}
\end{table}

\begin{table}
\centering
\small
\caption{Diehard tests results for 500 Mbit of true random data (S means ``SUCCESS'')}
\begin{tabular}{p{0.35\linewidth}ccc}
\hline
Statistical test   & P-Value 'A' & P-Value 'B' & Result\\
\hline
Birthday Spacing & 0.865773 & 0.842883  & S \\
Overlapping 5-permutation & 0.513342  & 0.471146 & S \\
Binary Rank for 31 x 31 matrices & 0.634721  & 0.755819 & S \\
Binary Rank for 32 x 32 matrices & 0.415482  & 0.352107 & S \\
Binary Rank for 8 x 8 matrices & 0.576012  &  0.361938  & S \\
Bitstream & 0.970227   & 0.051626 & S\\
Overlapping-Paris-Spares-Occupancy & 0.219893  & 0.015640  & S \\
Overlapping-Quardruples-Spares-Occupancy &  0.173054  &  0.015065 & S \\
DNA & 0.150832  &  0.016380 & S \\
Count-the-1's & 0.944274    & 0.058056 & S \\
Count-the-1's for specific bytes & 0.703417  & 0.365542 & S \\
Parking lot & 0.229559  &  0.239266 & S \\
Minimum distance & 0.794391  & 0.320832 & S \\
3D spheres & 0.448424 &  0.064255 & S \\
Squeeze & 0.317565  & 0.455628 & S \\
Overlapping sums & 0.250307  & 0.988284 & S \\
Runs & 0.814724  & 0.194590 & S \\
Craps & 0.840551  &  0.823860 & S \\
\hline
\end{tabular}
\label{tab:diehard}
\end{table}

Our chip achieved an extremely low energy per bit compared to the ones reported on standard CMOS technology (comparison with state of the art is reported in Table~\ref{tab:table1}), while the overall power at full speed was less than 499mW for the nominal 3.3V supply voltage at room temperature, making the chip suitable for low-power applications. The integration of the entropy source and the extractors enabled us to achieve the highest level of integration to date with negligible impact on the quality of the random sequence.\\
The micrograph of the QRNG chip is shown in Figure~\ref{fig:chip-micrograph}; the chip measures 10$\times$4.5mm$^2$ in this CMOS technology, while more advanced nodes would enable significant area reductions, provided similar noise, AP, and crosstalk performance in SPADs.

\begin{table*}[h!]
  \begin{center}
    \caption{Comparison Table. Comparison of the presented Quantum Random Number Generator with the state of the art}
    \label{tab:table1}
    
    \resizebox{\hsize}{!}{%
    \begin{tabular}{l|c|c|c|c|c|c|c|c|c|}
      \textbf{Performance} & \textbf{This Work} & \textbf{Stefanov\cite{stefanov2000optical}} & \textbf{Amri\cite{amri2016quantum}} & \textbf{Sanguinetti\cite{sanguinetti2014quantum,idq_chip}}  & \textbf{qStream\cite{quintessence}} & \textbf{Wei\cite{wei2011high}} & \textbf{Nie\cite{nie2014practical}} & \textbf{Dynes\cite{dynes2008high}} & \textbf{Matsumoto\cite{matsumoto20081200mum}}\\
      \hline
      \hline
      \textbf{Bit Rate} & 400Mb/s & 4-16Mb/s & 5-12Gb/s & 4.9Mb/s & 1Gb/s & 280Gb/s & 96Mb/s & 4Mb/s & 20Mb/s \\
      \hline
      \textbf{Thermal Noise} & $<0.1$\% & $<1$\% & $<1$\% & $<3$\% & Not specified & $<0.1$\% & $<0.1$\% & $<2$\% & N/A \\
      \hline
      \textbf{Full Integration} & yes & no & no & no & no & no & no & no & no \\
      \hline
      \textbf{On-chip Extractor} & yes & no & no & yes & yes & no & no & no & no \\
      \hline
      \textbf{Dimension ($mm^2$)} & 45 & 7728 & 25 & 22.5 & 10,892.4 & N/A & N/A & N/A & 0.012 \\
      \hline
      \textbf{Standard Tests/} & NIST, & NIST, DIEHARD, & \multirow{2}{*}{NIST} & NIST, ISTO/ & \multirow{2}{*}{NIST} & DIEHARD & \multirow{2}{*}{NIST} & NIST, & \multirow{2}{*}{NIST} \\
      \textbf{Certifications} & DIEHARD & METAL, CTL, AIS31 & & IEC-18031 & & STS & & DIEHARD & \\
      \hline
      \textbf{Power (mW)} & 499 & 1,500 & 25 & 83.44 & 12,780 & N/A & N/A & N/A & 1.9 \\
      \hline
      \textbf{FOM (nJ/bit)} & 1.25 & 93.7 & 0.0025 & 17 & 12.8 & N/A & N/A & N/A & 0.095 \\
      \hline
      \hline
      \end{tabular}
    }
  \end{center}
\end{table*}

\section{Conclusions}

We have presented the first integrated Quantum random number generator (QRNG) in a standard CMOS technology node. To demonstrate the possibility to integrate advanced functionalities with the entropy sources, we developed a modular architecture comprising an array of independently operating SPADs, illuminated by a DC-biased LED, and logic for random number extraction. The fabricated QRNG reaches a throughput of 400 Mbit/s with an extremely low energy per bit for the nominal 3.3~V supply voltage at room temperature. This characteristics make our chip suitable for portable low-power applications. Next steps will target the integration of the LED on the same substrate  and an efficient packaging of the whole chip to fulfill the compactness required by the System-in-Package (SiP) approach. The generation rate could also be improved by decreasing the number of SPADs and using the available space to implement a bigger extraction matrix.

\section{Acknowledgements} The work presented in this paper has been partially supported by the Eurostar framework (Progect Q-RANGER, Quantum - RAndom Number GEneRator). Davide Rusca thanks the EUs H2020 program under the Marie Sk\l{}odowska-Curie project QCALL (GA 675662) for financial support.

\bibliography{references} 
\bibliographystyle{IEEEtranS}

\section{Appendix: Entropy calculation}\label{app:model}

The model used in our paper follows the method introduced in Frauchiger et al.~\cite{Daniela}. We specify first the density operator $\rho$ corresponding to our QRNG. We consider our state as represented by different subsystems. A subsystem $I$ that, following a Poissonian distribution, encodes the number of photons. A subsystem $D$ that encodes the state of the detectors given by the random variable $S$. If $S = 1$ the detectors will experience a dark count and it will click independently of the absorption of a photon, otherwise the detector will behave normally. The string $\overline{s}$ corresponds to the state of each detector. Finally the subsystem $P$ considers the position of the photons with respect to the matrix of SPADs.

\begin{equation}
    \rho = \sum_{n,\overline{s}} P_N(n)P_S(\overline{s})\ket{n}\bra{n}_I\otimes\ket{\overline{s}}\bra{\overline{s}}_D\otimes\ket{\phi}\bra{\phi}_P^{\otimes n}
\end{equation}

The state of the $n$ photons is in the form $\ket{\phi}^{\otimes n}$. This corresponds to considering the photons as distinguishable and due to the fact that in beam-splitting experiment the behaviour of photons can be in principle described in this way as it is shown in the work of U.~Leonhardt~\cite{leonhardt2003}. Since the illumination of the SPAD matrix has been calibrated to be uniform, we can consider that each photon has equal probability to hit any detector. This means that we model the state $\ket{\phi}$ to be the $\ket{W}$ state of dimension $m$. If the illumination is imbalanced, then this state should be adjusted accordingly.

The positive-operator valued measure (POVM) can be described in terms of different operators for each detector $D_i$. The following expression describes the POVM's elements relative to the detection or no detection of a photon:

\begin{equation}
\begin{split}
    P^{n,1}_{D_i} := (\eta(\mathbb{1}-\ket{0}\bra{0}))^{\otimes n} \;\;\;\\
    P^{n,0}_{D_i} := ((1-\eta)\mathbb{1}+\eta\ket{0}\bra{0})^{\otimes n}
\end{split}
\end{equation}

For each photon arriving on the detector we consider a probability $\eta$ that the photon is absorbed by the detector. This information is considered to be unforeseeable by a third party. 

The operator related to generating a sequence $\overline{x}$ is now given by:

\begin{equation}
    \Pi^{\overline{x}} = \sum_{n,\overline{s}} \ket{n,\overline{s}}\bra{n,\overline{s}}_D \bigotimes_i P^{n,\overline{x}_i,\overline{s}_i}_i
\end{equation}
where the sum over $\overline{s}$ considers all the sequences consistent with the output sequence $\overline{x}$ (if $\overline{s}_i = 1$ this means $\overline{x}_i$ is necessarily $1$) and
\begin{equation}
  P^{n,\overline{x}_i,\overline{s}_i}_i=\begin{cases}
    P^{n,\overline{x}_i}, & \text{if $\overline{s}_i = 0$}\\
    id^{\otimes n}, & \text{if $\overline{s}_i = 1$}
  \end{cases}
\end{equation}
meaning that if the random variable $\overline{s}_i = 1$, the detector will click independently of the photon state $\ket{\phi}^{\otimes n}$.

Once the measurement operator and the preparation state of the device are defined we have to consider the classical noise. In order to model this we consider the following measurement:
\begin{equation}
    E^{n,\overline{s}} = \ket{n,\overline{s}}\bra{n,\overline{s}} \bigotimes_i \mathbb{1}^{\otimes n}
\end{equation}
the classical noise is defined by its outcome and corresponds to the random variable $N \in \{ 0, ..., \infty\}$ corresponding to the number of photon and $\overline{S}$ which is the sequence of random variables corresponding to the state of each detector with respect to a possible dark-count. 

The probability to have a certain output string of bits at each sampling event is given now by the Born rule:

\begin{equation}
    P(\overline{x}) = Tr(\Pi^{\overline{x}}\rho)
\end{equation}
which corresponds to the observed experimental distribution of the obtained bits strings.

However the probability distribution of importance in the presented model is given by the joint probability of the output variable and the classical noise, given again by the Born rule:

\begin{equation}
    P(\overline{x},n,\overline{s}) = Tr(\Pi^{\overline{x}}E^{n,\overline{s}}\rho(E^{n,\overline{s}})^\dag  )
\end{equation}

By simple combinatorial calculation it is possible to obtain the expression given in the main text:

\begin{equation}
\begin{split}
    P(\overline{x},n,\overline{s}) = P_N(n)P_S(\overline{s})\sum_{i=0}^{H(\overline{x})-H(\overline{s})} (-1)^i\\ 
        {{H(\overline{x})-H(\overline{s})}\choose{i}}
        \left(1-\eta-\frac{H(\overline{x})-i}{m}\eta\right)^n
\end{split}
\end{equation}

where the vector $\overline{s} = s_1, s_2, ... , s_m$ collects all the variables for dark counts for each detector, $H(\cdot)$ is the Hamming weight function and $m$ is the number of detector considered. This model corresponds to a generalization of the two model previously proposed for two detectors~\cite{Daniela}.

\end{document}